\documentclass[aps,reprint,superscriptaddress,longbibliography]{revtex4-2}
\usepackage[T1]{fontenc}
\usepackage{lmodern}
\usepackage{amsmath,amssymb,bm,graphicx,booktabs,float}
\usepackage{pgfplots}
\usepackage[hidelinks]{hyperref}
\newcommand{\dd}{\mathop{}\!\mathrm d}
\newcommand{\Tr}{\operatorname{Tr}}
\newcommand{\SU}{\mathrm{SU}}
\begin{document}
\title{Poles Are Not Enough: \\ \it{Zeros of Dyson--Schwinger Gluon Propagators in Yang--Mills Thermodynamics}}
\author{Marco Frasca}
\affiliation{Independent Researcher, Rome, Italy}
\author{Anish Ghoshal}
\affiliation{Department of Physics and Astronomy, University of Sussex, Brighton BN1 9RH, United Kingdom}
\author{Stefan Groote}
\affiliation{Institute of Physics, University of Tartu, W.~Ostwaldi 1, 50411 Tartu, Estonia}
\begin{abstract}
A normalized two-point function with several poles does not, on its own, fix the source-free thermal partition function. For the Gaussian measure of the original gluon field, a positive-residue rational propagator contributes determinants from its poles and from zeros interlaced between them. The zero factors distinguish this completion from a gas of independent pole fields and restore two ultraviolet polarizations per adjoint eigenvalue at every finite truncation. We apply the construction to a background-covariant completion of the elliptic Dyson--Schwinger gluon kernel. The first-order $\SU(3)$ transition is already present with a single pole, and the saturated six-pole result shifts $T_c/M_g$ by only $1.48\%$, where $T_c$ is the coexistence temperature and $M_g$ is the lowest pole mass. At six poles, the zero determinants raise $T_c/M_g$ by $2.31\%$ and lower $L/T_c^4$ by $12.4\%$, where $L$ is the latent heat and each prescription uses its own $T_c$. At fixed $M_g$ the dimensional latent heat decreases by $4.0\%$. The fixed-holonomy free-energy difference gives an ultraviolet test independent of the additive pressure normalization. At fixed pole number $P$, its asymptotic value agrees with the two-polarization Weiss result when the zeros are retained and is $(3P-1)/2$ times as large in the poles-only prescription.
\end{abstract}
\maketitle

\emph{\textbf{Introduction.}}---Infrared Yang--Mills correlators encode a dynamical scale, while the Polyakov loop characterizes thermal center symmetry~\cite{Polyakov,Susskind,Weiss,SvetitskyYaffe}. Lattice calculations establish the first-order $\SU(3)$ transition and benchmark its thermodynamics~\cite{Fukugita,Panero,Borsanyi,LuciniPanero}. Connecting the two needs more than a propagator. Normalized correlation functions lose the zero-source normalization that enters the free energy, so the fields and the functional measure used to realize a correlator become part of its thermal interpretation.

Functional and background-field approaches have shown that infrared gluon and ghost correlators generate center-sensitive Polyakov-loop potentials~\cite{Braun,BraunGroups,FukushimaKashiwa,Fischer}. Massive Landau--DeWitt calculations recover the expected second-order transition for $\SU(2)$ and first-order transition for $\SU(3)$~\cite{Reinosa,ReinosaTwo,Kondo}. Gribov-based thermodynamic constructions likewise retain nontrivial inverse-propagator structure, together with horizon conditions, condensates, or a temperature-dependent Gribov scale~\cite{ZwanzigerEOS,FukushimaSu,Canfora}. Here we ask a different question. If a multipole covariance is used as a Gaussian thermal kernel, which source-free determinant goes with the chosen field measure?

Fukushima and Kashiwa already retained a numerator-zero contribution in the thermal trace log of a Gribov--Stingl propagator [their Eq.~(8)]~\cite{FukushimaKashiwa}. Here we compare two realizations of the \emph{same normalized source covariance}. For finite rational propagators with distinct real poles and positive residues, the original-field Gaussian integral includes interlacing-zero factors, while the independent-field integral carries additional source-free determinants. The interlacing is a classical property of rational Stieltjes functions~\cite{Akhiezer}. This comparison identifies the determinant responsible for the extra species count in the independent-field interpretation of the elliptic Dyson--Schwinger tower~\cite{FGG}. It also yields an ultraviolet test at arbitrary finite pole number $P$ within the stated analytic class.

The elliptic Dyson--Schwinger (DS) solution supplies a concrete spectrum and relative residues~\cite{FrascaIR,FrascaMapping,FrascaQuantum,FrascaConfinement,FrascaGroote,FrascaPRD}. Its minimal background-covariant completion yields coexisting, locally stable center-symmetric and center-broken minima in the full $\SU(3)$ Cartan domain, a discontinuous background loop, and nonzero latent heat. The first-order transition itself is already present in the single-pole massive-background mechanism. We use it as a consistency check: the pole--zero completion keeps the expected phase structure but shifts its thermodynamic parameters at fixed truncation.

\begin{figure*}[t]
 \includegraphics[width=\textwidth]{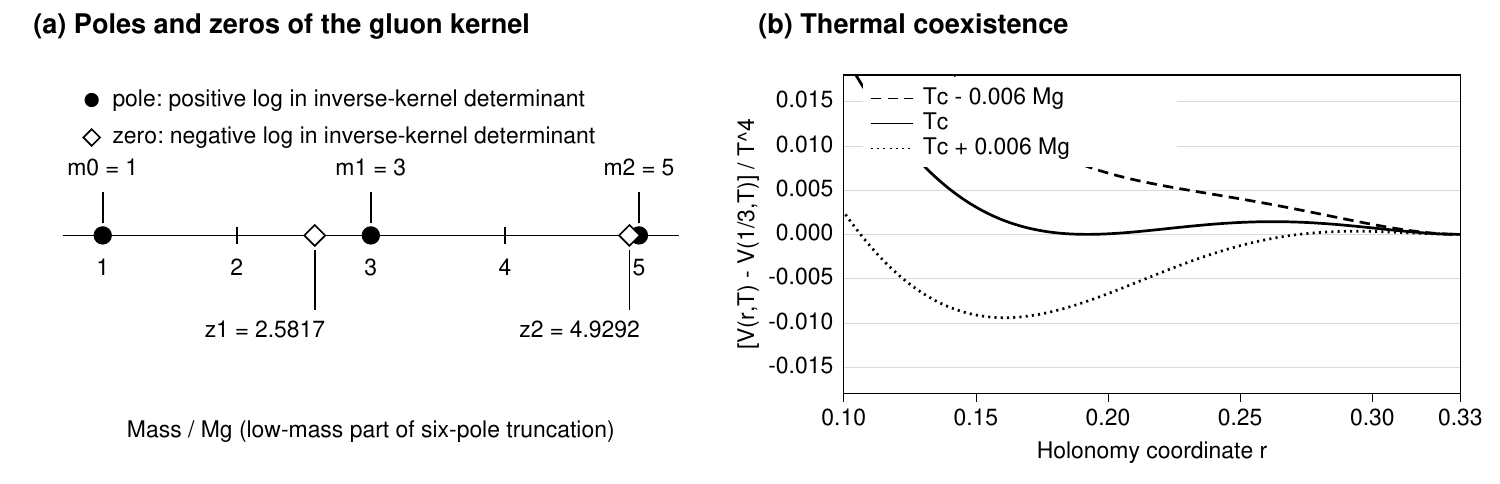}
 \caption{Pole--zero structure and its thermal application. (a) Low-mass part of the six-pole elliptic kernel: filled circles indicate poles and open diamonds indicate interlacing zeros. All five zero determinants enter the calculation; the three heaviest are omitted from this panel for clarity. (b) Six-pole $\SU(3)$ holonomy potential relative to the center-symmetric configuration, at $T_c/M_g=0.3690$ and nearby temperatures. Distinct minima coexist at $T_c$.}
 \label{fig:potential}
\end{figure*}

\emph{\textbf{Vacuum input and minimal thermal completion.}}---After Euclidean continuation, the scalar part of the ordinary Landau-gauge vacuum propagator is \cite{FrascaQuantum,FGG,FrascaPRD}
\begin{align}
 D(s)&=\sum_{j=0}^{\infty}\frac{B_j}{s+m_j^2},\qquad m_j=(2j+1)M_g,\label{eq:D}\\
 B_j&=\frac{\pi^3(2j+1)^2}{4K(i)^3}
 \frac{e^{-(j+1/2)\pi}}{1+e^{-(2j+1)\pi}},\qquad \sum_jB_j=1.\label{eq:B}
\end{align}
Here $K(i)\equiv K(m=-1)=1.311028777\ldots$. In the conventions of Ref.~\cite{FrascaPRD},
$M_g=\pi( Ng^2/2)^{1/4}\mu_0/[2K(i)]$, with $\mu_0$ a background integration constant. As in the cited construction, we take the branch with free massless ghosts and neglect the renormalized quantum mass shift. Here $N$ specifies the group $\SU(N)$, independently of the truncation order $P$. The scale $M_g=m_0$ is the lowest gluon pole mass of this DS branch. A physical glueball mass would instead require a gauge-invariant composite correlator.

For a static diagonal flat background, $\bar F_{\mu\nu}=0$, we write
\begin{equation}
 \begin{gathered}
 \bar A_0=\frac{2\pi T}{g}\operatorname{diag}(q_1,\ldots,q_N),\\
 \sum_iq_i=0,\qquad \ell(\bm q)=\frac1N\sum_i e^{2\pi iq_i}.
 \end{gathered}
 \label{eq:q}
\end{equation}
Although the connection is flat, the thermal holonomy can be nontrivial, since in general it cannot be removed by a strictly periodic gauge transformation. The vacuum input is in ordinary Landau gauge, $\partial_\mu A_\mu=0$.
For the holonomy calculation we split the total gauge field as $A_\mu=\bar A_\mu+a_\mu$, where $\bar A_\mu=\delta_{\mu0}\bar A_0$ is the prescribed background and $a_\mu$ is the quantum fluctuation integrated over in the Gaussian functional integral. We impose Landau--DeWitt gauge~\cite{Reinosa,ReinosaTwo},
\begin{equation}
 \bar D_\mu a_\mu=0,\qquad
 \bar D_\mu X=\partial_\mu X-ig[\bar A_\mu,X].
 \label{eq:LDW}
\end{equation}
At zero background this reduces to ordinary Landau gauge for the fluctuations. For the background-transverse fluctuations we assume the same scalar kernel. The replacement $D^{-1}(p^2)\mapsto D^{-1}(-\bar D^2)$ is therefore an assumption of the model; we do not derive it from the ordinary-gauge and background-field Hessians. With the convention in Eq.~\eqref{eq:LDW}, an adjoint weight $\lambda$ has frequency $2\pi T(n-\lambda\cdot\bm q)$. The final adjoint sum is unchanged under $\lambda\to-\lambda$. Masses and residues are held fixed, with equal electric and magnetic scalar kernels. Appendix~\ref{app:gauge} lists what this minimal completion does and does not fix.

This minimal completion recovers the vacuum covariance at zero background, is background covariant, respects center and Weyl symmetry as well as large-gauge periodicity, and, together with the gauge-sector determinant below, preserves the finite-truncation polarization count. This Gaussian model is the thermal completion we use throughout. Background-dependent vertices, electric-magnetic splitting, and interaction terms are discussed below, together with the scope of the model.

On a flat background the transverse projector has rank three. Gaussian integration over transverse fluctuations, the longitudinal constraint, and periodic Faddeev--Popov ghosts gives, as in standard background-field treatments~\cite{Braun,Reinosa},
\begin{equation}
 Z_G[\bm q]=\mathcal N(T)
 \frac{\det_{\rm adj}(-\bar D^2)^{1/2}}
 {\det_{\rm adj}[D^{-1}(-\bar D^2)]^{3/2}},
 \label{eq:ZG}
\end{equation}
Here $Z_G[\bm q]\equiv Z_G[J=0;\bm q]$ is the zero-source Gaussian partition function at fixed holonomy, with $J$ the gluon source. Appendix~\ref{app:completion} derives the ghost and longitudinal factors.
Determinants are regulated and made dimensionless with a common reference scale; gauge-volume and global zero modes are removed consistently. Their thermal background-independent normalization contributes $C(T)$ to $V_G=-(T/\mathcal V)\ln Z_G$, where $\mathcal V$ is spatial volume. For the specified quadratic model, Eq.~\eqref{eq:ZG} is exact. It says nothing about the interaction (skeleton) terms of the full Yang--Mills functional~\cite{CJT}.

This is where the present construction departs from Ref.~\cite{FGG}. A normalized Gaussian source functional fixes $D$ but leaves $Z_G[J=0;\bm q]$ undetermined. Independent fields with covariances $B_j/(s+m_j^2)$ reproduce the same source covariance when their sum couples to the source, but carry extra source-free determinants. These factors depend on temperature and, after covariantization, on holonomy. The two measures therefore define different Gaussian thermal theories with the same normalized source covariance; their determinant ratio is derived in Appendix~\ref{app:completion}. Equation~\eqref{eq:ZG} selects the original-field measure; its zero factors prevent counting its spectral poles as independent thermal species.

\emph{\textbf{Pole--zero thermal determinant.}}--- We keep $P$ distinct massive poles and rescale the retained residues according to
\begin{equation}
 b_j=\frac{B_j}{\sum_{k=0}^{P-1}B_k},\qquad
 \sum_{j=0}^{P-1}b_j=1.
 \label{eq:bj}
\end{equation}
Thus $D_P(s)=\sum_{j=0}^{P-1}b_j/(s+m_j^2)$ retains the ultraviolet behavior $D_P(s)\sim1/s$. This analytic class is what the selected Dyson--Schwinger branch provides; we do not propose it as a general spectral representation of the lattice Landau-gauge gluon propagator. The factorization below holds for any finite multipole covariance with positive residues and distinct real poles, whatever the mass spacing, and hence also for finite rational representations of other nonperturbative correlators. The normalization in Eq.~\eqref{eq:bj} makes the numerator monic. The resulting rational kernel has $P-1$ negative real zeros $-z_\alpha^2$, interlaced with its poles~\cite{Akhiezer} (a short proof and the six-pole zero positions are given in Appendix~\ref{app:zeros}):
\begin{equation}
 D_P(s)=\frac{\prod_{\alpha=1}^{P-1}(s+z_\alpha^2)}
 {\prod_{j=0}^{P-1}(s+m_j^2)},\qquad m_{\alpha-1}<z_\alpha<m_\alpha.
 \label{eq:factor}
\end{equation}
Consequently $\log D_P^{-1}$ is a sum of pole logarithms \emph{minus} zero logarithms. At the regulated determinant level this gives
\begin{equation}
 \det D_P^{-1}(-\bar D^2)=
 \frac{\prod_{j=0}^{P-1}\det(-\bar D^2+m_j^2)}
 {\prod_{\alpha=1}^{P-1}\det(-\bar D^2+z_\alpha^2)}.
 \label{eq:detfactor}
\end{equation}
A common regulator and subtraction convention are understood. In zeta-function regularization, factorizing products of massive Laplace-type operators may generate a multiplicative anomaly. For the flat background considered here this anomaly is local and independent of $T$ and $\bm q$, so it is absorbed into the vacuum normalization $C(T)$ and does not affect the holonomy-dependent thermal results. No independent fields are assigned to the zero factors; they enter the denominator of $\det D_P^{-1}$ and therefore contribute with the opposite sign to the pole factors in $\log D_P^{-1}$ and in the thermal free energy. Equation~\eqref{eq:detfactor} is understood up to this local term. Note that it differs from the residue-weighted prescription $\sum_j b_j\log(s+m_j^2)$: the normalized residues fix the positions of the zeros, not thermodynamic degeneracies. The interlacing is illustrated in Fig.~\ref{fig:potential}(a).

We define the thermal contribution of one real scalar per adjoint eigenvalue
\begin{equation}
 f(x,\theta)=\frac{1}{2\pi^2}\int_0^\infty\dd y\,y^2
 \operatorname{Re}\ln(1-e^{-\varepsilon+i\theta}).\label{eq:f}
\end{equation}
Here $x=m/T$, $y=|\bm p|/T$, and $\varepsilon=\sqrt{x^2+y^2}$ are the dimensionless mass, spatial momentum, and energy, respectively. The real part implements the pairing of adjoint weights $\lambda$ and $-\lambda$. Its massless limit, the Weiss potential~\cite{Weiss}, is
\begin{equation*}
 f(0,\theta)=-\frac1{\pi^2}\sum_{n=1}^{\infty}\frac{\cos(n\theta)}{n^4}.
\end{equation*}
For $x>0$ an equivalent representation (see Appendix~\ref{app:zeros}) is $f=-x^2\sum_{n\ge1}K_2(nx)\cos(n\theta)/(2\pi^2n^2)$, where $K_2(z)$ is the modified Bessel function of the second kind. The Matsubara sum in Eq.~\eqref{eq:ZG} yields the holonomy-dependent thermal free-energy density as
\begin{align}
 u=\frac{V_G(\bm q,T)-C(T)}{T^4}
 &=\sum_{\lambda\in{\rm adj}}F_P(2\pi\lambda\cdot\bm q;t),\label{eq:V}\\
 F_P(\theta;t)&=3\sum_{j=0}^{P-1}f(m_j/T,\theta)\nonumber\\
 &\quad-3\sum_{\alpha=1}^{P-1}f(z_\alpha/T,\theta)-f(0,\theta).
 \nonumber
\end{align}
We use $t=T/M_g$, so that $m_j/T=(2j+1)/t$ and $z_\alpha/T=(z_\alpha/M_g)/t$. Here \(C(T)\) denotes a holonomy-independent contribution common to both phases; it cancels from their free-energy difference but contributes to the absolute pressure and entropy. The quantities \(z_\alpha>0\), indexed by \(\alpha=1,\ldots,P-1\), locate the zeros of the truncated propagator, \(D_P(-z_\alpha^2)=0\). For positive residues, they interlace with the pole masses, \(m_{\alpha-1}<z_\alpha<m_\alpha\), and are determined by the masses and residues; they do not represent additional particle species. The adjoint sum includes all oriented roots and $N-1$ neutral weights. Center symmetry, Weyl symmetry, and large-gauge periodicity are manifest.

For $T\ll M_g$, where $M_g$ is the lowest pole scale of the selected propagator branch, the massive determinants are suppressed and the remaining inverted massless potential favors center symmetry. More generally, at fixed finite $P$, $T\to\infty$ gives $3P-3(P-1)-1=2$ massless polarizations per adjoint eigenvalue. This count relies on the zero determinants; keeping only the pole determinants of the transverse kernel would give $3P-1$. At the center-broken minimum the original-field completion reproduces $(N^2-1)\pi^2/45$ for $p/T^4$ if $C/T^4\to0$. This ultraviolet result applies to the Gaussian model at fixed finite $P$. Likewise, the low-temperature absolute pressure need not be that of a physical glueball gas~\cite{Trotti}.

\emph{\textbf{Physical application: $\SU(3)$ coexistence.}}---A full fundamental alcove is parametrized by the gaps $a=q_2-q_1$, $b=q_3-q_2$, with $a,b\ge0$ and $a+b\le1$. Equation~\eqref{eq:V} becomes
\begin{equation}
 \begin{aligned}
 u(a,b;t)=2[&F_P(0;t)+F_P(2\pi a;t)\\
 &+F_P(2\pi b;t)+F_P(2\pi(a+b);t)],
 \end{aligned}
 \label{eq:alcove}
\end{equation}
with $t=T/M_g$ as before.
For $a=b=r$, one has $\bm q=(-r,0,r)$ and $\ell=(1+2\cos 2\pi r)/3$ the corresponding normalized traced Polyakov loop. The center-symmetric point is $r_c=1/3$; the center-broken minimum near coexistence need not occur at $r=0$.

For the six-pole truncation, after the convergence checks below, we obtain
\begin{align}
 t_c&=0.3690, & r_d&=0.1921,\label{eq:result}\\
 \Delta\ell&=0.5705, & \frac{L}{T_c^4}&=0.4680.\nonumber
\end{align}
Here $\Delta\ell$ is the jump of the background-field order parameter $\ell(\bar A_0)$ in this approximation; it should not be identified with the renormalized expectation value of the composite Polyakov loop; its meaning in a finite-volume, center-symmetric ensemble is discussed in Appendix~\ref{app:cartan}. As noted above, $C(T)$ drops out of the latent heat. Numerical errors are below the quoted digits.
Figure~\ref{fig:potential}(b) shows the two degenerate minima at coexistence, separated along the real holonomy path by a barrier of free-energy density $\Delta V_{\mathrm{bar}}=1.44\times10^{-3}T_c^4$.

To test transverse stability we use $\bm q=(-r+\sigma,-2\sigma,r+\sigma)$. In terms of the alcove gaps, $a=r-3\sigma$ and $b=r+3\sigma$, so $\sigma\to-\sigma$ exchanges $a$ and $b$. Charge conjugation followed by the corresponding eigenvalue permutation is a symmetry of $u$, and the mixed Hessian entry therefore vanishes identically at $\sigma=0$. The Hessians of the dimensionless potential $u$ of Eq.~\eqref{eq:V} are diagonal by symmetry and, at coexistence, are
\begin{equation}
 \begin{aligned}
 \mathcal H_c&\simeq\operatorname{diag}(2.135,6.404),\\
 \mathcal H_d&\simeq\operatorname{diag}(2.229,29.57).
 \end{aligned}
 \label{eq:H}
\end{equation}
Here $\mathcal H_{ij}=\partial_{\xi_i}\partial_{\xi_j}u$ with $(\xi_1,\xi_2)=(r,\sigma)$.
At the center-symmetric point, $Z_3$ symmetry makes the Hessian isotropic in orthonormal Cartan coordinates. Since $|\partial\bm q/\partial r|^2=2$ and $|\partial\bm q/\partial\sigma|^2=6$, it follows exactly that $(\mathcal H_c)_{\sigma\sigma}/(\mathcal H_c)_{rr}=3$, which the quoted values confirm. Both minima are locally stable in the full Cartan sector (finite-difference checks are given in Appendix~\ref{app:cartan}). A 7381-point triangular scan, with spacing $1/120$ over the complete alcove, at $t_c$ and $t_c\pm0.005$ finds no competing phase below these branches and their center images.

The thermodynamic discontinuity follows directly from the background effective potential. In the thermodynamic limit, where the holonomy integral is dominated by its saddle points (Appendix~\ref{app:cartan}), the equilibrium free-energy density is
\begin{equation}
 f_{\rm eq}(T)=\min_{\bm q\in\mathcal A}V_G(\bm q,T),
 \label{eq:feq}
\end{equation}
where $\mathcal A$ is the fundamental alcove. At coexistence the confined and deconfined minima have equal free energies. If $\Delta u=u_d-u_c$, then $\Delta u(t_c)=0$, and the envelope theorem removes derivatives of the minimizing coordinates. Consequently,
\begin{equation}
 \frac{L}{T_c^4}=\frac{e_d-e_c}{T_c^4}
 =-t_c\left.\frac{\partial\Delta u}{\partial t}\right|_{t_c}>0,
 \label{eq:latent}
\end{equation}
where $e_i=V_i-T\,\dd V_i/\dd T$ is the energy density of phase $i$. The term $4T^3\Delta u$ vanishes at $T_c$, while stationarity eliminates the implicit $t$ dependence of the minimizing holonomies. Equal free energies with unequal slopes give a kink in the free energy. Together with $\Delta\ell\ne0$ and the positive Hessians, this makes the transition first order within the present closure.

\emph{\textbf{Convergence and interpretation.}}---For $P=1,2,4,6$, we find $t_c=0.363636$, $0.368909$, $0.369021$, and $0.369021$. The single-pole case reproduces the known first-order mechanism of Ref.~\cite{Reinosa} (see also Appendix~\ref{app:analytic}); adding the higher poles together with their zeros shifts the scale by $1.48\%$. If instead the zeros are dropped at fixed $P=6$, we find $t_c=0.3607$ and $L/T_c^4=0.5341$. Including them raises $t_c$ by $2.31\%$. Using the respective transition temperatures, $L/T_c^4$ decreases by $12.4\%$. At fixed $M_g$, $L/M_g^4=t_c^4(L/T_c^4)$ decreases from $0.009041$ to $0.008678$, a $4.0\%$ reduction. Both prescriptions give first-order coexistence (Table~\ref{tab:comparison} in Appendix~\ref{app:eos}). Near coexistence the heavier masses are thermally suppressed, and the six-pole determinant is numerically saturated; increasing quadrature from 160 to 240 nodes changes $t_c$ by less than $10^{-12}$.

No polynomial barrier~\cite{Pisarski,Meisinger} or thermal transition scale~\cite{FGG} is fitted. The kernel and the measure fix the dimensionless transition, while an absolute $T_c$ needs an independent vacuum calibration of $M_g$.

\begin{figure}[t]
 \centering
 \begingroup
 \pgfplotsset{compat=1.17}
 \begin{tikzpicture}
 \begin{semilogxaxis}[
  width=\columnwidth,height=0.68\columnwidth,
  xmin=0.3,xmax=100,ymin=-0.6,ymax=9.1,
  xtick={0.3,0.5,1,10,100},xticklabels={0.3,0.5,1,10,100},
  ytick={-0.5,0,1,3,6,8.5},
  xlabel={$T/T_{c,\mathrm{ref}}$},
  ylabel={$\mathcal D_X/(16\pi^2/81)$},
  tick label style={font=\footnotesize},
  label style={font=\small},
  legend style={font=\footnotesize,draw=none,fill=none,at={(0.035,0.96)},anchor=north west},
  legend cell align=left,
 ]
 \addplot[black,thin,forget plot] coordinates {(0.3,0) (100,0)};
 \addplot[black,thick,solid] coordinates {
  (0.30,-0.4965299) (0.32,-0.4943913) (0.34,-0.4914681)
  (0.35,-0.4896791) (0.370363,-0.4853062) (0.39191,-0.4795517) (0.414711,-0.4721479) (0.438839,-0.4628232)
  (0.464371,-0.4513137) (0.491387,-0.4373759) (0.519976,-0.4208002) (0.550228,-0.4014227) (0.58224,-0.3791359)
  (0.616114,-0.3538977) (0.651959,-0.3257371) (0.68989,-0.2947567) (0.730027,-0.2611318) (0.7725,-0.2251063)
  (0.817444,-0.1869844) (0.865002,-0.1471199) (0.915327,-0.1059026) (0.968581,-0.0637431) (1.02493,-0.0210570)
  (1.08456,0.0217511) (1.14766,0.0643018) (1.21443,0.1062534) (1.28509,0.1473111) (1.35985,0.1872333)
  (1.43897,0.2258345) (1.52269,0.2629846) (1.61128,0.2986061) (1.70502,0.3326685) (1.80422,0.3651812)
  (1.90918,0.3961856) (2.02026,0.4257474) (2.1378,0.4539487) (2.26217,0.4808805) (2.39378,0.5066372)
  (2.53305,0.5313113) (2.68042,0.5549898) (2.83637,0.5777515) (3.00139,0.5996650) (3.17601,0.6207883)
  (3.36079,0.6411687) (3.55631,0.6608431) (3.76322,0.6798390) (3.98216,0.6981754) (4.21384,0.7158648)
  (4.459,0.7329140) (4.71842,0.7493258) (4.99294,0.7651000) (5.28342,0.7802351) (5.59081,0.7947291)
  (5.91608,0.8085805) (6.26027,0.8217887) (6.62449,0.8343553) (7.0099,0.8462835) (7.41773,0.8575792)
  (7.84929,0.8682508) (8.30596,0.8783090) (8.7892,0.8877672) (9.30055,0.8966408) (9.84165,0.9049472)
  (10.4142,0.9127058) (11.0201,0.9199373) (11.6613,0.9266636) (12.3397,0.9329075) (13.0576,0.9386927)
  (13.8173,0.9440427) (14.6212,0.9489817) (15.4719,0.9535335) (16.372,0.9577215) (17.3245,0.9615689)
  (18.3324,0.9650981) (19.399,0.9683309) (20.5276,0.9712881) (21.7219,0.9739898) (22.9857,0.9764551)
  (24.323,0.9787019) (25.7381,0.9807475) (27.2355,0.9826079) (28.8201,0.9842981) (30.4968,0.9858324)
  (32.2711,0.9872238) (34.1486,0.9884846) (36.1353,0.9896262) (38.2377,0.9906590) (40.4623,0.9915927)
  (42.8164,0.9924363) (45.3074,0.9931980) (47.9434,0.9938853) (50.7327,0.9945051) (53.6843,0.9950637)
  (56.8076,0.9955670) (60.1127,0.9960202) (63.61,0.9964281) (67.3108,0.9967951) (71.2269,0.9971251)
  (75.3708,0.9974217) (79.7559,0.9976883) (84.396,0.9979277) (89.3061,0.9981428) (94.5019,0.9983358)
  (100,0.9985091)
 };
 \addlegendentry{Pole--zero}
 \addplot[black,thick,densely dashed] coordinates {
  (0.30,-0.4965298) (0.32,-0.4943912) (0.34,-0.4914679)
  (0.35,-0.4896789) (0.370363,-0.4853057) (0.39191,-0.4795504) (0.414711,-0.4721447) (0.438839,-0.4628157)
  (0.464371,-0.4512970) (0.491387,-0.4373407) (0.519976,-0.4207290) (0.550228,-0.4012845) (0.58224,-0.3788785)
  (0.616114,-0.3534362) (0.651959,-0.3249385) (0.68989,-0.2934208) (0.730027,-0.2589673) (0.7725,-0.2217029)
  (0.817444,-0.1817826) (0.865002,-0.1393783) (0.915327,-0.0946664) (0.968581,-0.0478141) (1.02493,0.0010316)
  (1.08456,0.0517542) (1.14766,0.1042748) (1.21443,0.1585568) (1.28509,0.2146081) (1.35985,0.2724807)
  (1.43897,0.3322689) (1.52269,0.3941057) (1.61128,0.4581586) (1.70502,0.5246258) (1.80422,0.5937311)
  (1.90918,0.6657206) (2.02026,0.7408588) (2.1378,0.8194264) (2.26217,0.9017173) (2.39378,0.9880369)
  (2.53305,1.0786999) (2.68042,1.1740276) (2.83637,1.2743449) (3.00139,1.3799762) (3.17601,1.4912392)
  (3.36079,1.6084378) (3.55631,1.7318527) (3.76322,1.8617308) (3.98216,1.9982719) (4.21384,2.1416163)
  (4.459,2.2918309) (4.71842,2.4488963) (4.99294,2.6126958) (5.28342,2.7830069) (5.59081,2.9594958)
  (5.91608,3.1417162) (6.26027,3.3291120) (6.62449,3.5210248) (7.0099,3.7167045) (7.41773,3.9153241)
  (7.84929,4.1159967) (8.30596,4.3177949) (8.7892,4.5197702) (9.30055,4.7209735) (9.84165,4.9204740)
  (10.4142,5.1173767) (11.0201,5.3108382) (11.6613,5.5000795) (12.3397,5.6843964) (13.0576,5.8631672)
  (13.8173,6.0358575) (14.6212,6.2020223) (15.4719,6.3613062) (16.372,6.5134414) (17.3245,6.6582435)
  (18.3324,6.7956068) (19.399,6.9254982) (20.5276,7.0479499) (21.7219,7.1630525) (22.9857,7.2709473)
  (24.323,7.3718187) (25.7381,7.4658874) (27.2355,7.5534028) (28.8201,7.6346373) (30.4968,7.7098797)
  (32.2711,7.7794302) (34.1486,7.8435953) (36.1353,7.9026840) (38.2377,7.9570040) (40.4623,8.0068586)
  (42.8164,8.0525444) (45.3074,8.0943488) (47.9434,8.1325489) (50.7327,8.1674103) (53.6843,8.1991855)
  (56.8076,8.2281144) (60.1127,8.2544232) (63.61,8.2783245) (67.3108,8.3000175) (71.2269,8.3196882)
  (75.3708,8.3375098) (79.7559,8.3536430) (84.396,8.3682364) (89.3061,8.3814275) (94.5019,8.3933430)
  (100,8.4040992)
 };
 \addlegendentry{Poles only}
 \addplot[black,thick,densely dotted] coordinates {
  (0.30,0.0023134) (0.32,0.0037392) (0.34,0.0056880)
  (0.35,0.0068807) (0.370363,0.0097962) (0.39191,0.0136331) (0.414711,0.0185702) (0.438839,0.0247895)
  (0.464371,0.0324686) (0.491387,0.0417728) (0.519976,0.0528473) (0.550228,0.0658103) (0.58224,0.0807476)
  (0.616114,0.0977092) (0.651959,0.1167076) (0.68989,0.1377195) (0.730027,0.1606885) (0.7725,0.1855314)
  (0.817444,0.2121449) (0.865002,0.2404145) (0.915327,0.2702224) (0.968581,0.3014573) (1.02493,0.3340210)
  (1.08456,0.3678361) (1.14766,0.4028498) (1.21443,0.4390379) (1.28509,0.4764054) (1.35985,0.5149872)
  (1.43897,0.5548459) (1.52269,0.5960704) (1.61128,0.6387724) (1.70502,0.6830839) (1.80422,0.7291541)
  (1.90918,0.7771470) (2.02026,0.8272392) (2.1378,0.8796176) (2.26217,0.9344782) (2.39378,0.9920246)
  (2.53305,1.0524666) (2.68042,1.1160184) (2.83637,1.1828966) (3.00139,1.2533175) (3.17601,1.3274928)
  (3.36079,1.4056252) (3.55631,1.4879018) (3.76322,1.5744872) (3.98216,1.6655146) (4.21384,1.7610776)
  (4.459,1.8612206) (4.71842,1.9659309) (4.99294,2.0751306) (5.28342,2.1886713) (5.59081,2.3063305)
  (5.91608,2.4278108) (6.26027,2.5527414) (6.62449,2.6806832) (7.0099,2.8111364) (7.41773,2.9435494)
  (7.84929,3.0773312) (8.30596,3.2118633) (8.7892,3.3465135) (9.30055,3.4806490) (9.84165,3.6136493)
  (10.4142,3.7449178) (11.0201,3.8738921) (11.6613,4.0000530) (12.3397,4.1229309) (13.0576,4.2421115)
  (13.8173,4.3572383) (14.6212,4.4680149) (15.4719,4.5742042) (16.372,4.6756276) (17.3245,4.7721623)
  (18.3324,4.8637379) (19.399,4.9503321) (20.5276,5.0319666) (21.7219,5.1087017) (22.9857,5.1806315)
  (24.323,5.2478791) (25.7381,5.3105916) (27.2355,5.3689352) (28.8201,5.4230916) (30.4968,5.4732532)
  (32.2711,5.5196201) (34.1486,5.5623968) (36.1353,5.6017893) (38.2377,5.6380027) (40.4623,5.6712391)
  (42.8164,5.7016962) (45.3074,5.7295658) (47.9434,5.7550326) (50.7327,5.7782735) (53.6843,5.7994570)
  (56.8076,5.8187430) (60.1127,5.8362821) (63.61,5.8522163) (67.3108,5.8666783) (71.2269,5.8797922)
  (75.3708,5.8916732) (79.7559,5.9024286) (84.396,5.9121576) (89.3061,5.9209517) (94.5019,5.9288953)
  (100,5.9360661)
 };
 \addlegendentry{Independent gas}
 \end{semilogxaxis}
 \end{tikzpicture}
 \endgroup
 \caption{Free-energy difference between center-symmetric and trivial holonomy at six poles. The scale $T_{c,\mathrm{ref}}=0.3690212583M_g$ is common to all curves. The vertical normalization is the two-polarization Weiss limit. The thin horizontal line marks $\mathcal D_X=0$. The pole--zero and poles-only curves begin below it, and the independent gas stays above it for $T>0$. The high-temperature limits are $1$, $8.5$, and $6$, respectively. Each curve is independent of its own additive $C_X(T)$.}
 \label{fig:fixedholonomy}
\end{figure}
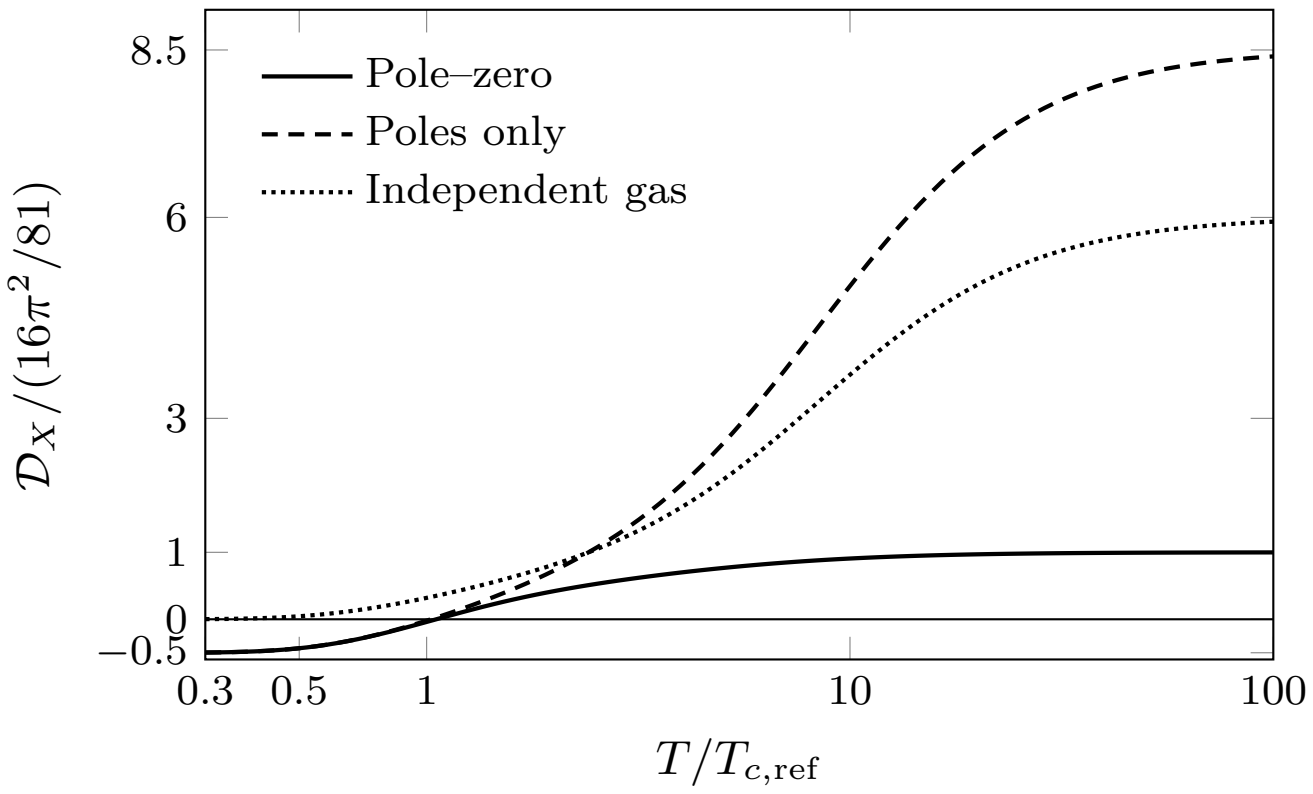

\emph{\textbf{Ultraviolet test at fixed holonomy.}} The absolute pressure depends on the additive normalization $C_X(T)$ of each prescription $X$. The dimensionless difference
\begin{equation}
 \mathcal D_X(T)=
 \frac{V_X(\bm q_{\rm cs},T)-V_X(\bm 0,T)}{T^4}
 \label{eq:fixedholonomy}
\end{equation}
cancels $C_X(T)$ separately for each $X$. Here $\bm q_{\rm cs}=(-1/3,0,1/3)$ is the center-symmetric background and $\bm 0$ is trivial holonomy. The neutral weights cancel, and the six roots give $\mathcal D_X=6[F_P^X(2\pi/3;t)-F_P^X(0;t)]$. From the massless scalar result, $6[f(0,2\pi/3)-f(0,0)]=8\pi^2/81$. Thus at fixed $P$, $\mathcal D_X\to 8\pi^2n_X/81$, with $n_X=2$, $3P-1$, and $2P$ for the pole--zero, poles-only, and independent-gas prescriptions, respectively. Figure~\ref{fig:fixedholonomy} shows the six-pole results. At low temperature, $\mathcal D_X<0$ for the pole--zero and poles-only prescriptions, so the center-symmetric background has lower free energy than trivial holonomy. Both curves change sign near $T_{c,\mathrm{ref}}$. The independent gas has $\mathcal D_X>0$ at every $T>0$ and favors trivial adjoint holonomy throughout.

\emph{\textbf{Equilibrium Gaussian thermodynamics.}} With $C_X(T)=0$, the six-pole pressures at $10T_{c,\mathrm{ref}}$ are $p_{G,X}/T^4=1.586$, $8.658$, and $6.357$ for the three prescriptions (Fig.~\ref{fig:eqeos} in Appendix~\ref{app:eos}). The poles-only pressure is about $5.5$ times the pole--zero pressure in this convention. On the pole--zero confined branch, $p_G/T_c^4\simeq-0.002075$ at coexistence and $s_G/T_c^3\simeq-0.171$ on approach from below, where $s_G=\dd p_G/\dd T$. A constant vacuum subtraction leaves the entropy unchanged. A common temperature-dependent $C(T)$ shifts the entropy by $-C'(T)$ and cancels from the latent heat. Related one-loop artifacts in the massive background-field approach were removed at two-loop order~\cite{ReinosaTwo}. The full Gaussian-sector curves and their normalization are given in Appendix~\ref{app:eos}.

The scope of the calculation is the Gaussian sector of a fixed vacuum kernel in background Landau--DeWitt gauge. The positive-residue input does not reproduce the Schwinger-function positivity violation of the lattice Landau-gauge gluon propagator~\cite{CucchieriPositivity,PositivityLattice} (Appendix~\ref{app:analytic}); complex-conjugate-pole and Gribov-type kernels are the natural next extension~\cite{ZwanzigerEOS,FukushimaSu,Canfora}. For comparison, our $L/T_c^4=0.468$ lies well below the continuum lattice values $1.117(40)$ and $1.349(38)$ at spatial aspect ratios 8 and 6~\cite{Latent2021}. We also note that background covariance alone does not guarantee independence of the fluctuation gauge fixing. A quantitative comparison would require temperature-dependent electric and magnetic kernels, background vertices, interaction terms~\cite{ReinosaTwo}, and the corresponding gauge identities. The common term $C(T)$ affects absolute thermodynamics but cancels from phase differences; bubble nucleation would in addition need gradient terms.

\emph{\textbf{Conclusion.}}---The poles of a multipole propagator do not, by themselves, fix its thermodynamics. For the Gaussian measure of the original field, the inverse-kernel determinant contains pole factors and interlacing-zero factors. The zeros are not extra excitations; they encode the difference between the original-field measure and a collection of independent pole fields. Retaining them prevents the ultraviolet species count from growing with the truncation and preserves two gluon polarizations per adjoint eigenvalue at fixed finite order.

The elliptic Dyson--Schwinger kernel provides an application of the measure comparison. Its minimal background Landau--DeWitt completion preserves the first-order $\SU(3)$ transition found at one pole. At six poles the zero factors raise $T_c/M_g$ by $2.31\%$ and reduce $L$ by $4.0\%$ at fixed $M_g$. The much larger ultraviolet change is visible in the fixed-holonomy free-energy difference. Its poles-only limit at six poles is $8.5$ times the two-polarization Weiss value, independently of the additive normalization.

The distinction between measures holds for any finite positive-residue rational kernel, independently of the elliptic spectrum. Natural next steps are to extend it to positivity-violating complex-pole and Gribov-type propagators~\cite{CucchieriPositivity,Canfora}, and to match the Gaussian completion to thermal kernels, background vertices, and interactions, as needed for quantitative Yang--Mills thermodynamics.
\par\smallskip
\noindent\textit{Data availability.}---No external datasets enter the numerical calculation. The Python scripts and numerical outputs needed to reproduce the pole locations, thermal integrals, convergence tests, Hessians, and figures are supplied on reasonable request to the authors.
\par\smallskip
\noindent\textit{Acknowledgments.}---A.G. acknowledges support from the Royal Society, United Kingdom, under grant NIF\ R1\ 253963. Generative AI tools were used for literature review, certain calculations and language editing and presentation; all derivations, numerical calculations, interpretations, and final text were checked and approved by the authors.

\appendix

\section{Definition of the thermal completion}
\label{app:completion}
\noindent Here we keep apart the exact Gaussian identities and the assumptions needed to extend the elliptic DS covariance to finite temperature. The main text takes the massive scalar propagator and the massless ghost branch as input, uses the small-mass-shift approximation, and covariantizes the vacuum kernel on a flat temporal background. This procedure does not solve the thermal hierarchy, nor does it reconstruct its vertices.

For a real Gaussian field $A$ with Euclidean propagator (covariance) $D$ and fixed flat integration measure, Gaussian integration gives
\begin{equation}
 \int \mathcal DA\,e^{-\frac12 A D^{-1}A+JA}
 =\mathcal C\,(\det D^{-1})^{-1/2}e^{\frac12 JDJ}.
 \label{eq:gaussian}
\end{equation}
The sign of the source term is a convention and does not affect the argument. Knowing only the normalized functional $Z[J]/Z[0]$ leaves the zero-source normalization undetermined, together with its temperature and background dependence, which is exactly what thermodynamics needs.

This point is relevant for the auxiliary-field representation used in our earlier thermal analysis of Ref.~\cite{FGG}. More generally, let independent real variables
$A_j$ have covariance $b_j/(s+m_j^2)$ and let their sum couple to $J$. Their normalized Gaussian integral has covariance
\begin{equation}
 D_P(s)=\sum_{j=0}^{P-1}\frac{b_j}{s+m_j^2}.
\end{equation}
However, the unnormalized auxiliary integral has the determinant
\begin{equation}
 Z_{\rm aux}[0]\propto\prod_{j=0}^{P-1}\left[\det\frac{s+m_j^2}{b_j}\right]^{-1/2},
\end{equation}
whereas the original-field integral has $Z_A[0]\propto[\det D_P^{-1}(s)]^{-1/2}$. For normalized residues, $\sum_jb_j=1$, and a consistent subtraction of field-independent constants,
\begin{equation}
 \frac{Z_{\rm aux}[0]}{Z_A[0]}
 \propto
 \left[
  \det\!\left(\prod_{\alpha=1}^{P-1}(s+z_\alpha^2)\right)
 \right]^{-1/2}.
 \label{eq:ratio}
\end{equation}
Here $z_\alpha>0$ denotes the mass scale associated with the propagator zero at $s=-z_\alpha^2$, with $\alpha=1,\ldots,P-1$. This factor changes the thermodynamics: the two constructions share the normalized source covariance but are different Gaussian theories. The same identity follows from changing variables to the sum $S=\sum_j A_j$ and $P-1$ orthogonal auxiliary coordinates and integrating out the latter, which carry exactly the extra normalization. In the main text we use Eq.~\eqref{eq:gaussian} for the original gluon variable, so the model is not equivalent to a gas of $P$ independent physical particles. This is a choice of thermal completion; fixing a unique Yang--Mills normalization would require matching it to the underlying theory.

In the background Landau constraint the transverse fluctuation has three Euclidean components. For each nonzero momentum, integrating the longitudinal mode with $\delta(\bar D\cdot a)$, where $a_\mu=A_\mu-\bar A_\mu$ is the quantum fluctuation and the delta functional imposes the Landau--DeWitt constraint, contributes $[\det(-\bar D^2)]^{-1/2}$. The complex periodic ghost contributes $\det(-\bar D^2)$. Together they give the positive half-power of the massless determinant in the numerator of the partition function, Eq.~\eqref{eq:ZG}. Residual global zero modes are removed with the gauge-group volume. Tensor covariantization beyond a flat background, interaction vertices, and fluctuation--background identities would require further input.

\section{Minimal completion and gauge distinction}
\label{app:gauge}
\noindent
We write $\bar D_\mu X=\partial_\mu X-ig[\bar A_\mu,X]$ and impose $\bar D_\mu a_\mu=0$. The vacuum covariance is supplied in ordinary Landau gauge; using its scalar kernel for background-transverse fluctuations is an additional assumption. Background-gauge covariance is not the same as independence of the fluctuation gauge fixing, and we have not established the latter for this closure. The construction is minimal in that it uses only the vacuum scalar kernel, the covariant Laplacian on a flat holonomy, and the stated gauge and ghost measure. It respects background covariance, Weyl and center symmetry, and periodicity under large gauge transformations. The finite-truncation ultraviolet count is a further consistency check of the combined determinant. These conditions do not single out a unique completion, since background-dependent vertices, operators that vanish at zero background, electric--magnetic thermal splitting, and interaction functionals are not constrained by the vacuum two-point function. Once the finite kernel and the Gaussian measure are fixed, the pole--zero identity is exact; how it matches Yang--Mills thermodynamics is a separate physical question.

\section{Zeros and the thermal determinant}
\label{app:zeros}
\noindent For positive retained $b_j$, define $g(x)=D_P(-x)$. Between neighboring poles,
\begin{equation}
 g'(x)=\sum_j\frac{b_j}{(m_j^2-x)^2}>0.
\end{equation}
On $(m_{j-1}^2,m_j^2)$, $g$ increases from $-\infty$ to $+\infty$; it therefore has exactly one zero. The numerator of the rational function has degree $P-1$, so these account for all its zeros. The normalization $\sum_jb_j=1$ fixes its leading coefficient. This proves the finite-tower factorization used in the main text. The negative logarithms of the zeros are algebraic factors of the determinant; they are not negative-norm particle states, and no independent fields are assigned to them.

For six poles, masses are $1,3,5,7,9,11$ in units of $M_g$ and the zero masses are approximately 
\begin{equation}
 \begin{aligned}
 (z_1,z_2,z_3)/M_g&=(2.58169820,\,4.92919228,\,6.99109031),\\
 (z_4,z_5)/M_g&=(8.99914448,\,10.99993098).
 \end{aligned}
\end{equation}
The unnormalized retained residue sum is $\sum_{j=0}^{5}B_j=0.9999991650$. Rescaling this sum changes only the overall normalization of the kernel and leaves the zeros where they are. The source-normalized covariance and the source-free normalization still have to be kept apart, as discussed above.

The thermal part of $\frac12\Tr\ln(s+m^2)$ per adjoint eigenvalue follows by summing frequencies with twist $\theta$:
\begin{align}
 &\frac{T}{2}\int\frac{\dd^3p}{(2\pi)^3}
 \ln\!\left[
  1-2e^{-\beta\omega_p}\cos\theta+e^{-2\beta\omega_p}
 \right]\nonumber\\
 &\qquad=T^4 f(m/T,\theta),
 \qquad \omega_p=\sqrt{p^2+m^2}.
\end{align}
Expanding the logarithm and performing the radial integral yields the Bessel series for $f(x,\theta)$ in the main text. At $x=0$, it is evaluated analytically with
\begin{equation}
 f(0,2\pi a)=\frac{\pi^2}{3}B_4(\{a\}),\qquad
 B_4(a)=a^4-2a^3+a^2-\frac1{30}.
\end{equation}
In particular, $f(0,0)=-\pi^2/90$ fixes the real-scalar normalization. At any fixed finite $P$ the ultraviolet determinant count is two polarizations. We do not interchange the limits $P\to\infty$ and $T\to\infty$, and we assess numerical convergence only in the transition region.

\section{The full Cartan sector and the thermodynamic limit}
\label{app:cartan}
\noindent Ordered eigenvalues in the fundamental alcove can be written
\begin{equation}
 \begin{gathered}
 \bm q=\left(-\frac{2a+b}{3},\frac{a-b}{3},\frac{a+2b}{3}\right),\\
 a\ge0,\quad b\ge0,\quad a+b\le1.
 \end{gathered}
\end{equation}
The finite-dimensional Haar factor is proportional to
\begin{equation}
 J_H(a,b)=\sin^2(\pi a)\sin^2(\pi b)\sin^2[\pi(a+b)].
\end{equation}
It is nonzero at both interior coexisting minima. It enters as a global integration measure rather than as an extensive local potential. The integral over homogeneous holonomies is part of the model; including inhomogeneous holonomy fluctuations would go beyond it.

Near a nondegenerate minimum $q_i$, the two-dimensional saddle point contribution is proportional to
\begin{equation}
 J_H(q_i)\frac{2\pi T}{\mathcal V\sqrt{\det V_G''(q_i)}}
 e^{-\mathcal V V_G(q_i,T)/T}.
\end{equation}
The measure and saddle prefactors therefore contribute only $\mathcal O(\ln\mathcal V/\mathcal V)$ to the free-energy density. At coexistence the confined saddle and three center-related deconfined saddles contribute. A center-symmetric finite-volume ensemble has vanishing average traced loop; the jump quoted in the main text refers to a selected deconfined sector, or equivalently the magnitude at a dominant saddle, with the usual thermodynamic symmetry-breaking prescription.

Write $V_i=C+T^4u_i(T/M_g)$. Stationarity eliminates derivatives of the minimizing coordinates in $\dd V_i/\dd T$. At coexistence $u_d=u_c$, so
\begin{equation}
 L=e_d-e_c=-T_c\frac{\dd(V_d-V_c)}{\dd T}
 =-T_c^4 t_c\,\partial_t(u_d-u_c).
\end{equation}
Thus the nonzero calculated slope difference gives a cusp in the minimum free energy as $\mathcal V\to\infty$. A common $C(T)$ cancels from this relation but remains essential for absolute thermodynamics. The phase ordering near coexistence, which we verify numerically, supports this saddle-point picture, although we have no analytic proof of global uniqueness.

The coordinates used for the stability test are $a=r-3\sigma$, $b=r+3\sigma$. Symmetry gives zero mixed Hessian at $\sigma=0$. Finite differences with step $10^{-4}$ give
\begin{equation}
 \begin{aligned}
 \mathcal H_c&=
 \begin{pmatrix}
  2.1346244 & 0\\
  0 & 6.4038740
 \end{pmatrix},\\[3pt]
 \mathcal H_d&=
 \begin{pmatrix}
  2.2289935 & 0\\
  0 & 29.5663539
 \end{pmatrix}.
 \end{aligned}
\end{equation}
Reducing the step to $5\times10^{-5}$ changes entries by at most $1.4\times10^{-6}$ and $2.4\times10^{-5}$, respectively. The individual Hessian entries depend on the coordinates, but positive definiteness at a stationary point does not. They should be read as curvatures of the potential, not as gradient-normalized screening masses.

\section{Equilibrium Gaussian-sector equation of state}
\label{app:eos}
We compare three fully specified finite-$P$ prescriptions at fixed vacuum masses. Write $s=-\bar D^2$ and use the same adjoint holonomy coupling. The original-field and poles-only determinants are
\begin{align}
 Z_{\mathrm{pz}}&\propto(\det_{\rm adj}s)^{1/2}
 \left[\frac{\prod_\alpha\det_{\rm adj}(s+z_\alpha^2)}{\prod_j\det_{\rm adj}(s+m_j^2)}\right]^{3/2},\\
 Z_{\mathrm{po}}&\propto(\det_{\rm adj}s)^{1/2}
 \prod_j[\det_{\rm adj}(s+m_j^2)]^{-3/2}.
\end{align}
Both retain a single shared ghost--longitudinal sector. Three independent transverse components per pole with this same shared sector give exactly $Z_{\mathrm{po}}$. An auxiliary-field realization with the same multiplicities and background coupling therefore does not produce a fourth curve.

The independent gas actually used in Eqs.~(19)--(20) of Ref.~\cite{FGG} instead assigns $n_g=2(N^2-1)$ real scalar copies to each pole and contains no residual massless determinant. For normalized retained residues $b_j$, at zero holonomy it is
\begin{equation}
 Z_{\mathrm{ind}}[0]\propto
 \prod_{j=0}^{P-1}\left[\det\frac{-\partial^2+m_j^2}{b_j}\right]^{-(N^2-1)}.
\end{equation}
To compare at a common background we specify its adjoint-covariant extension
\begin{equation}
 Z_{\mathrm{ind}}[\bm q]\propto
 \prod_j\left[\det_{\rm adj}\frac{s+m_j^2}{b_j}\right]^{-1}.
\end{equation}
This extension is our own additional choice; Ref.~\cite{FGG} did not compute a holonomy potential. Subtracting the field-independent measure constants consistently, the thermal functions are
\begin{align}
 F_P^{\mathrm{pz}}(\theta;t)&=3\sum_j f(m_j/T,\theta)\nonumber\\
 &\quad-3\sum_\alpha f(z_\alpha/T,\theta)-f(0,\theta),\\
 F_P^{\mathrm{po}}(\theta;t)&=3\sum_j f(m_j/T,\theta)-f(0,\theta),\\
 F_P^{\mathrm{ind}}(\theta;t)&=2\sum_j f(m_j/T,\theta).
\end{align}
Residues fix the source covariance and the zero locations, not thermal multiplicities. The auxiliary measure with three transverse components and a shared ghost sector discussed above gives $F_P^{\mathrm{aux,shared}}=F_P^{\mathrm{po}}$; the third curve here uses the original FGG multiplicity and therefore differs from it. Because $f(x,\theta)\ge f(x,0)$, the positive independent-gas sum is minimized at a center element, with trivial adjoint holonomy, at every $T>0$. Its equilibrium pressure is then just the holonomy-free gas pressure, and with these fixed masses there is no center-restoring transition and no latent heat.

\begin{table}[t]

\caption{Same-truncation comparison on the real holonomy path. Both Cartan curvatures are positive at coexistence. The poles-only prescription retains the same ghost--longitudinal term but omits all zero determinants.}
\begin{ruledtabular}
\begin{tabular}{rlrrr}
$P$ & Prescription & $t_c$ & $\Delta\ell$ & $L/T_c^4$\\
1 & pole-zero & 0.363636 & 0.571948 & 0.505298\\
1 & poles-only & 0.363636 & 0.571948 & 0.505298\\
2 & pole-zero & 0.368909 & 0.570554 & 0.468700\\
2 & poles-only & 0.360718 & 0.572695 & 0.533760\\
6 & pole-zero & 0.369021 & 0.570524 & 0.467960\\
6 & poles-only & 0.360696 & 0.572700 & 0.534147\\
\end{tabular}
\end{ruledtabular}
\label{tab:comparison}
\end{table}

Table~\ref{tab:comparison} lists the coexistence comparison; the independent gas is treated separately above. At $P=6$, including zeros raises $t_c$ by $2.31\%$ and lowers $L/T_c^4$ by $12.4\%$ relative to poles-only. Here $L/T_c^4$ uses the respective transition temperatures. At a common $M_g$, $L/M_g^4$ is $0.009041$ for poles only and $0.008678$ with zeros, giving a $4.0\%$ decrease in $L$. The $1.48\%$ change in $t_c$ from one to six poles includes both the higher poles and their zeros.

\begin{figure*}[t]
 \centering\includegraphics[width=\textwidth]{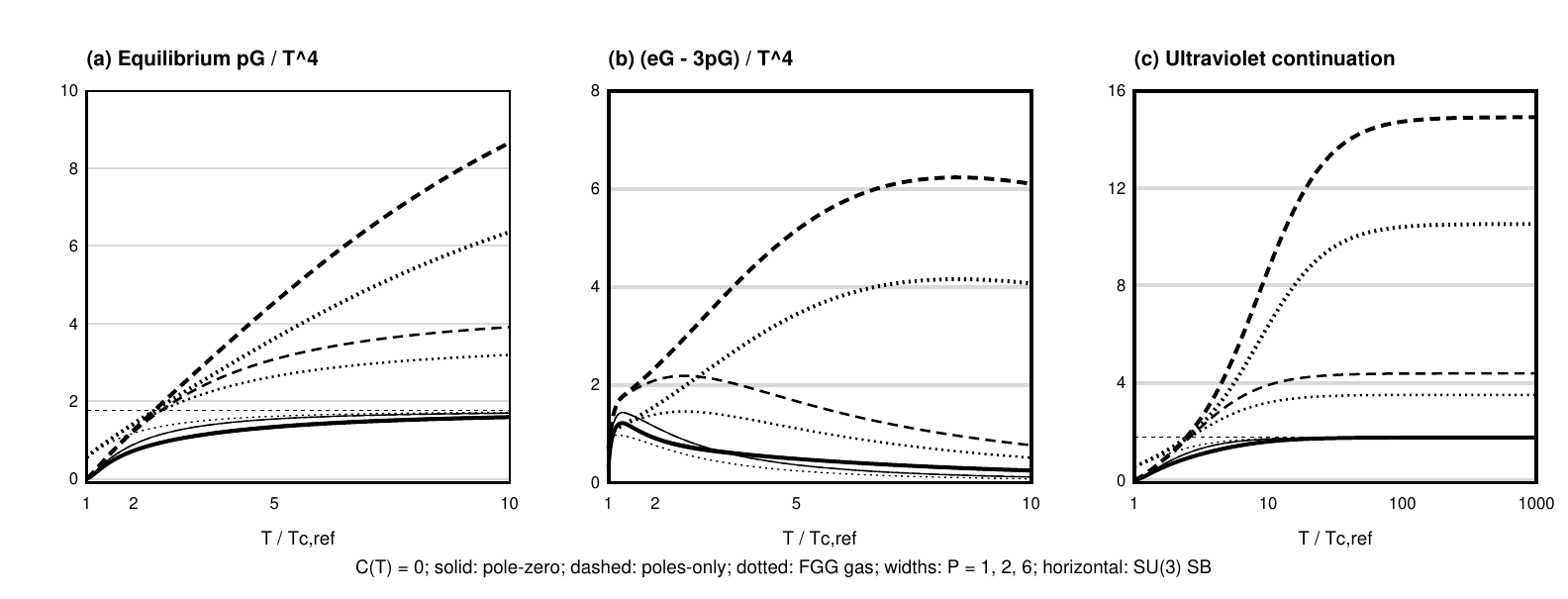}
 \caption{Equilibrium Gaussian-sector thermodynamics with additional $C(T)=0$, retaining neutral thermal determinants. Solid, dashed, and dotted curves denote pole--zero, poles-only, and the independent FGG-multiplicity gas, respectively; increasing width denotes $P=1,2,6$. All curves use the common temperature $T/T_{c,\mathrm{ref}}$. (a) Pressure and (b) interaction measure up to $10T_{c,\mathrm{ref}}$. (c) Continuation to $1000T_{c,\mathrm{ref}}$ to resolve the fixed-truncation ultraviolet limits. The horizontal reference is $p_{\mathrm{SB}}/T^4=8\pi^2/45$. The independent gas has no thermal center transition; its specified adjoint extension is minimized at trivial adjoint holonomy.}
 \label{fig:eqeos}
\end{figure*}

\paragraph{Normalization and equilibrium.}
For each prescription $X$, we retain the complete thermal determinant, including the $N-1$ neutral adjoint weights and its holonomy-independent thermal terms. Only the additional normalization is set to $C_X(T)=0$; no subtraction at a reference temperature, $T$-dependent pressure adjustment, or residue weighting is made. Thus
\begin{align}
 u_X(a,b;t)&=2\{F_P^X(0;t)+F_P^X(2\pi a;t)\nonumber\\
 &\qquad+F_P^X(2\pi b;t)+F_P^X(2\pi(a+b);t)\},\\
 \bar u_X(t)&=\min_{a,b\ge0,\ a+b\le1}u_X(a,b;t),\\
 \frac{p_{G,X}}{T^4}&=-\bar u_X(t),\qquad
 \frac{e_{G,X}-3p_{G,X}}{T^4}=-t\frac{d\bar u_X}{dt}.
\end{align}
The envelope theorem evaluates the temperature derivative at the stationary holonomy; the massless term has no explicit $t$ derivative. At coexistence the derivative is one-sided, and the jump equals $L/T_c^4$. The common horizontal coordinate in Fig.~\ref{fig:eqeos} is $T/T_{c,\mathrm{ref}}$, with $T_{c,\mathrm{ref}}=0.3690212583M_g$, the six-pole pole--zero critical temperature. This is a common reference, not each curve's own $T_c$; the independent gas has none. At the reference coexistence point the plotted pole--zero value is the deconfined one-sided limit.

The curves show Gaussian-sector pressures and should not be read as a quantitative Yang--Mills equation of state. A nonzero additional $C_X(T)$ would shift $p/T^4$ by $-C_X/T^4$ and the interaction measure by $(4C_X-TC'_X)/T^4$. Differences at the same physical temperature cancel a common $C(T)$; a subtraction at one reference temperature cannot remove an arbitrary function. For $P=6$ and $C=0$, the pole--zero prescription gives $p_G/T_c^4\simeq-0.002075$ at coexistence. From $s_G=\dd p_G/\dd T$, the one-sided entropy on the confined branch is $s_G(T_c^-)/T_c^3=-4\bar u_X(t_c)-t_c\bar u_X'(t_c^-)\simeq-0.171$. The negative slope persists immediately below $T_c$ on this branch. A constant shift of the vacuum energy leaves the entropy unchanged, while a common $C(T)$ shifts it by $-C'(T)$ and cancels from the latent heat. Negative pressure and entropy also occur at one loop in the massive background-field approach, where two-loop corrections yield positive values~\cite{ReinosaTwo}.

\paragraph{Ultraviolet test.}
At fixed finite $P$, the three polarization counts are
\begin{equation}
 n_{\mathrm{pz}}=2,\qquad n_{\mathrm{po}}=3P-1,\qquad n_{\mathrm{ind}}=2P.
\end{equation}
Their asymptotic pressures are $(N^2-1)\pi^2 n_X/90$, while each interaction measure tends to zero for fixed masses. Thus the pole--zero pressure approaches the same two-polarization limit at every truncation, whereas both independent-pole prescriptions acquire a growing species count. This is a finite-$P$ statement; the $P\to\infty$ and $T\to\infty$ limits are not interchanged.

For the fixed-holonomy test in Eq.~\eqref{eq:fixedholonomy}, the center-symmetric adjoint phases consist of two neutral phases and six phases congruent to $\pm 2\pi/3$. The evenness and periodicity of $F_P^X$ yield the sixfold scalar difference quoted in the main text. Its massless value follows from $B_4(1/3)-B_4(0)=4/81$. The curves in Fig.~\ref{fig:fixedholonomy} use the Bessel series at these two holonomies, with the massless term evaluated analytically. This difference can be evaluated at any temperature without tracking the equilibrium saddle.

At $10T_{c,\mathrm{ref}}$ the six-pole masses are not all thermally negligible: $m_5/T=2.98$. The pressures are $1.58577$, $8.65775$, and $6.35670$ for pole--zero, poles-only, and independent gas. At $1000T_{c,\mathrm{ref}}$ they are $1.75457$, $14.9120$, and $10.5262$, close to $1.75460$, $14.9141$, and $10.5276$, respectively. At $10T_{c,\mathrm{ref}}$, the three pressures are well separated and remain below their respective six-pole asymptotes. We use the high-temperature continuation only to count the degrees of freedom of the fixed vacuum kernel; it is not meant as a physical extrapolation of that kernel to arbitrarily high temperature.

\paragraph{Numerical checks and relation to previous thermodynamics.}
Our script \texttt{equilibrium\_eos.py} computes the pressure, stationary holonomy, and analytic temperature derivative. It searches the real branch and the full triangular alcove with a 48-subdivision grid and a two-dimensional local refinement. At selected temperatures, 96 subdivisions and 240 momentum nodes are compared with the baseline 160 nodes. The largest pressure change is below $5\times10^{-9}$; a centered derivative of the equilibrium pressure agrees with the directly integrated interaction measure to $2\times10^{-7}$. These searches are numerical checks and do not classify all stationary points analytically. The independent-gas minimum follows analytically from positivity of its coefficients. CSV data retain the common $T/M_g$, prescription, pole count, minimizing gaps, pressure, and interaction measure.

For the direct derivative, with $\varepsilon=\sqrt{y^2+x^2}$ and $R=1-2e^{-\varepsilon}\cos\theta+e^{-2\varepsilon}$,
\begin{equation}
 x\partial_x f(x,\theta)=\frac{x^2}{2\pi^2}\int_0^\infty\!dy\,
 \frac{y^2 e^{-\varepsilon}(\cos\theta-e^{-\varepsilon})}{\varepsilon R}.
\end{equation}
Summing this expression with the same determinant coefficients gives $-t\partial_tu_X$; the massless term contributes zero. Independent-gas derivatives use the coefficient 2 per pole.

Thermodynamics with Gribov-based dispersions, nonlocal kernels, and specified auxiliary sectors has been studied previously~\cite{ZwanzigerEOS,FukushimaSu,FlorkowskiTransport}. Fukushima and Kashiwa~\cite{FukushimaKashiwa} explicitly include the numerator zero of a rational Gribov--Stingl fit in the thermal trace log. The comparison here uses two field realizations of the same positive-residue covariance and tracks the source-free factor that distinguishes them. Our comparison isolates the measure-dependent species count, using the same pole masses and common physical temperatures, without identifying the analytic structures or causal prescriptions of the different kernels.

\section{Analytic structure and relation to other kernels}
\label{app:analytic}
\noindent For the positive-residue scalar kernel, the zero-spatial-momentum Schwinger function is
\begin{equation}
 \Delta(\tau)=
 \sum_j\frac{B_j}{2m_j}e^{-m_j|\tau|}>0.
\end{equation} This property differs from the positivity-violating lattice gluon Schwinger function; it is a limitation of the adopted vacuum branch as a complete Landau-gauge correlator. It does not by itself establish physical colored asymptotic states, which require the full gauge theory.

The original covariant Gribov form $D_{\rm G}(s)=s/(s^2+\gamma^4)$ has $D_{\rm G}(0)=0$ and complex-conjugate poles, unlike the finite nonzero infrared propagator observed in four-dimensional Landau-gauge simulations~\cite{DudalOliveira}. Refined Gribov--Zwanziger kernels include additional condensate terms and can have $D(0)>0$, so this objection does not carry over to them. Nor do we take complex poles or reflection-positivity violation as evidence of real-time acausality, which would require a continuation prescription and an analysis of causal correlators. The real-pole interlacing argument used here makes no assumption about such a continuation.

The massive background calculation of Ref.~\cite{Reinosa} introduces an explicit mass parameter in its gauge-fixed action and calibrates it from vacuum correlators; it does not derive the elliptic DS spectrum used here. Its one-pole potential is nevertheless the natural comparison for the thermal mechanism. In our case the DS input fixes the relative masses and residues, while the overall scale $M_g$ and the thermal completion are separate inputs.

\end{document}